\documentclass[twocolumn]{aastex631}

\usepackage{academicons}
\usepackage{amsmath}
\usepackage{hyperref}
\usepackage{natbib}
\usepackage{xcolor}
\usepackage{xspace}

\newcommand{\e}{{\langle}e{\rangle}}

\newcommand{\Me}{\ensuremath{M_{\oplus}}\xspace} 
\renewcommand{\Re}{\ensuremath{R_{\oplus}}\xspace}

\newcommand{\Mjup}{\ensuremath{M_{\mathrm{Jup}}}\xspace}

\shorttitle{CLS Giant Planet and Brown Dwarf Eccentricities}
\shortauthors{Gilbert et al.}

\begin{document}

\title{Orbital Eccentricities Suggest a Gradual Transition from Giant Planets to Brown Dwarfs}

%% The \author command takes an optional argument which is the 16 digit ORCID. 
%% The syntax is: \author[xxxx-xxxx-xxxx-xxxx]{Author Name}
\correspondingauthor{Gregory J. Gilbert}
\email{ggilbert@caltech.edu}

\author[0000-0003-0742-1660]{Gregory J. Gilbert}
\affiliation{Department of Physics \& Astronomy, University of California Los Angeles, Los Angeles, CA 90095, USA}
\affiliation{Department of Astronomy, California Institute of Technology, Pasadena, CA 91125, USA}

\author[0000-0002-4290-6826]{Judah Van Zandt}
\altaffiliation{NASA FINESST Fellow}
\affiliation{Department of Physics \& Astronomy, University of California Los Angeles, Los Angeles, CA 90095, USA}

\author[0000-0003-0967-2893]{Erik A. Petigura}
\affiliation{Department of Physics \& Astronomy, University of California Los Angeles, Los Angeles, CA 90095, USA}

\author[0000-0002-8965-3969]{Steven Giacalone}
\altaffiliation{NSF Astronomy and Astrophysics Postdoctoral Fellow}
\affiliation{Department of Astronomy, California Institute of Technology, Pasadena, CA 91125, USA}

\author[0000-0001-8638-0320]{Andrew W. Howard}
\affiliation{Department of Astronomy, California Institute of Technology, Pasadena, CA 91125, USA}

\author[0000-0002-9305-5101]{Luke B. Handley}
\affiliation{Department of Astronomy, California Institute of Technology, Pasadena, CA 91125, USA}

\begin{abstract}

To date, hundreds of sub-stellar objects with masses between $1-80\ \Mjup$ have been detected orbiting main-sequence stars. The current convention uses the deuterium-burning limit, $M_c \approx 13 \Mjup$ to divide this population between giant planets and brown dwarfs. However, this classification heuristic is largely divorced from any formation physics and may not accurately reflect the astrophysical nature of these objects. Previous work has suggested that a transition from ``planet-like'' to ``brown-dwarf-like'' characteristics occurs somewhere in the range $1-10 \Mjup$, but precise the crossover mass and whether the transition is gradual or abrupt remains unknown. Here, we explore how the occurrence rate, host star metallicity, and orbital eccentricities vary as a function of mass in a sample of 70 Doppler-detected sub-stellar objects ($0.8 < M_c/\Mjup < 80$) from the California Legacy Survey. Our population consists of objects near and beyond the water ice line ($1 < a / \text{AU} < 10$), providing valuable clues to the details of giant planet formation physics at a location in the proto-stellar disk where planet formation efficiency is thought to be enhanced. We find that occurrence rate, host star metallicity, and orbital eccentricity all change gradually across the mass range under consideration, suggesting that ``bottom-up'' core accretion mechanisms and ``top-down'' gravitational instability mechanisms produce objects that overlap in mass. The observed eccentricity distributions could arise either from different formation channels or from post-formation dynamical interactions between massive sub-stellar objects. 

\end{abstract}

%\keywords{Exoplanets (498) --- Bayesian statistics (1900) --- Astronomy data modeling (1859)}

\section{Introduction}
\label{sec:intro}

Astronomers have detected hundreds of sub-stellar objects with masses between $1-80\ \Mjup$ orbiting main-sequence stars. Objects below this mass limit ($M_c < 1\ \Mjup$) are classified as planets and are thought to form via rapid oligarchic growth of solid cores in the proto-stellar disk, followed by a period of runaway H/He gas accretion \citep{Pollack1996}. Objects above this mass limit ($M_c > 80 \ \Mjup$) form via direct gravitational collapse of molecular clouds and become main sequence stars which fuse hydrogen in their cores \citep[e.g.,][]{McKeeOstriker2007}. Objects in the intermediate regime ($M_c \approx 1-80 \, \Mjup$) may form through bottom-up core accretion \citep{Pollack1996} or through top-down gravitational instability \citep{Boss1997}. Historically, the deuterium-burning limit, $M_c \approx 13 \, \Mjup$ \citep{Spiegel2011}, has been used to divide the intermediate-mass population between giant planets and brown dwarfs. However, while deuterium burning can alter cooling rates and the age-luminosity relationship, its influence on bulk mass is far less pronounced \citep{Chabrier2014}. Furthermore, it is likely that the mass domains of the largest giant planets and the smallest brown dwarfs overlap considerably. Consequently, there exists an order-of-magnitude mass range over which sub-stellar objects cannot yet be unambiguously classified. The aim of this study is to clarify this distinction.

Numerous prior studies have sought to identify the transition from bottom-up formation to top-down formation on the basis of host star metallicity. Giant planets have long been known to form predominantly around high-metallicity stars \citep{FischerValenti2005}, a correlation that does not hold for small planets \citep{Buchhave2012} or for high-mass brown dwarfs \citep{MaGe2014}. This correlation implies that enhanced disk metallicities are required for massive planetary cores to form and trigger runaway accretion prior to gas disk dispersal \citep{IdaLin2004-metallicity}. In contrast, gravitational instability has no such metallicity requirement \citep{Chabrier2014}. Using various populations of sub-stellar companions, some studies have identified a host star metallicity transition around $4-10 \, \Mjup$ \citep{Santos2017, Schlaufman2018, Adibekyan2019}, while others have identified a transition around $35-55 \, \Mjup$ \citep{MaGe2014, MataSanchez2014, MaldonadoVillaver2017}. It is possible that the lower mass threshold corresponds to a transition from core accretion to disk instability, while the higher mass threshold corresponds to a transition from disk instability to direct fragmentation and gravitational collapse of the molecular cloud \citep{Jappsen2005, ForganRice2011}.

Orbital eccentricity provides a complementary and powerful probe of giant planet and brown dwarf formation physics. Studies of Doppler-detected, directly imaged, and transiting sub-stellar objects have consistently identified a qualitative trend wherein lower mass ``planet-like'' objects exhibit lower eccentricities compared to their higher mass ``brown-dwarf-like'' counterparts. Based on observations of 396 Doppler-detected giant planets (median $P=382$ days), \citet{Kipping2013} recovered an eccentricity distribution which was well-described by a beta distribution peaked at $e=0$ with mean eccentricity $\e \approx 0.23$. In contrast, an analysis of Doppler-detected brown dwarf companions to FGK stars ($P \lesssim 2000$ days) by \cite{MaGe2014} observed a much more uniform distribution of eccentricities between $e\approx 0.0-0.9$. Similarly, \citet{Bowler2020} found that directly imaged giant planets ($M_p < 15 \, \Mjup$) at separations of 5--100 AU exhibit a preference for low eccentricities ($e \approx 0.05-0.25$), whereas their directly imaged brown dwarf counterparts ($M_p > 15 \, \Mjup$) possess a wide range of eccentricities peaked near $e \approx 0.6$. Although the exact values for the directly imaged population are somewhat sensitive to modeling choices \citep{Nagpal2023, DoO2023}, the broad conclusion that giant planets exhibit a narrow range of low eccentricities compared to a broad, high eccentricity distribution for brown dwarfs is robust. The trend toward low eccentricities for giant planets also holds true for close-in transiting planets ($P < 100$ days), which \citet{Gilbert2025} observed to possess a mean eccentricity $\e \approx 0.2$, consistent with both the intermediate-period Doppler-detected population \citep{Kipping2013} and the long-period directly-imaged population \citep{Bowler2020}. Thus, eccentricity measurements across a wide range of star-companion separations are all consistent with a paradigm whereby giant planets form by core accretion \citep{Ida2013-eccentricity} and brown dwarfs form by gravitational instability \citep{Vorobyov2013, MaGe2014}.

Here, we present evidence that the transition from planet-like objects to brown-dwarf-like objects near the ice line occurs gradually between ${\sim}2-13 \Mjup$. Our conclusion is based on a hierarchical Bayesian analysis of the orbital eccentricities of 70 Doppler-detected sub-stellar objects from the California Legacy Survey \citep[CLS;][]{Rosenthal2021}. We combined CLS radial velocity measurements with \textit{Hipparcos-Gaia} astrometry, thereby breaking the usual inclination degeneracy inherent to Doppler-only studies, allowing us to measure true mass $M_c$ as opposed to minimum mass $M_c \sin i$. Leveraging the homogeneous search strategy employed by CLS, we corrected for survey completeness to recover the unbiased, intrinsic eccentricity distribution of objects between $0.8-80 \Mjup$ at orbital separations between $1-10$ AU. Two companion papers to this study (Van Zandt et al., in review, Giacalone et al., in review) investigated the companion occurrence rates and host star metallicity properties of the CLS sample, reaching similar conclusions.

This paper is organized as follows. In \S2 we review the California Legacy Survey. In \S3 we describe our hierarchical analysis and discuss demographic trends in orbital eccentricities as a function of companion mass and separation. In \S4 we discuss the astrophysical implications of our findings, and in \S5 we summarize our conclusions.

\section{Observations}\label{sec:observations}

\subsection{California-Legacy Survey radial velocities}

The California Legacy Survey \citep[CLS;][]{Rosenthal2021} monitored 719 bright, nearby stars taking over 100,000 precision radial velocity (RV) measurements using the Keck-HIRES, APF-Levy, and Lick-Hamilton spectrographs over more than thirty years. CLS recovered 226 companions, of which, by the classical dividing lines, 34 were stellar ($M \sin i > 80 ~\Mjup$), 19 were brown dwarfs ($13 < M \sin i / \Mjup \leq 80$), and 173 --- including 14 new discoveries --- were planetary ($M \sin i \le 13 ~\Mjup$). These detections spanned star-planet separations out to 20 AU and masses $M_p$ down to ${\sim}0.01~\Mjup$. In addition, CLS derived uniform properties for host stars using high-resolution spectral matching \citep[\texttt{SpecMatch};][]{Petigura2015, Yee2017} and isochrone fitting \citep[\texttt{Isoclassify;}][]{Huber2017}. The homogeneous observing strategy of CLS has made it a valuable resource for probing exoplanet demographics, particularly for planets at or near the water ice line \citep[e.g.][]{Fulton2021, VanZandtPetigura2024}.

For our present study, we chose to focus on the region of parameter space between $a = 1-10$ AU and $M_c = 0.8 - 80 \Mjup$ (Figure 1). We chose this mass interval to cover Jovian planets, super-Jovians, and brown dwarfs, and we chose 1--10 AU in order to limit our analysis separations over which CLS had high sensitivity to companions and full orbital phase coverage. A total of 70 out of 226 companions reside in this mass-separation interval.

\subsection{Hipparcos-Gaia astrometry}\label{subsec:astrometry}

Doppler measurements alone provide only a minimum companion mass $M_c \sin i$ rather than a true mass $M_c$ because radial velocities cannot constrain orbital inclination. To break this degeneracy and recover true masses, we combined CLS Doppler measurements with the \textit{Hipparcos-Gaia} catalog of astrometric accelerations \citep[HGCA;][]{Brandt2021}. For a more detailed description of this procedure, see Van Zandt et al., in review. A summary is presented below.

HGCA provides astrometric proper motions for a catalog of over 115,000 stars, determined by aligning \textit{Hipparcos} astrometric observations \citep{HipparcosCatalog} with those of \textit{Gaia} EDR3 \citep{GaiaEDR3}. By combining these astrometric accelerations with RV measurements, one may infer the presence of massive long-period planets and place separate constraints on $M_c$ and $\sin i$.

We used \texttt{Orvara} \citep{Orvara2021} to uniformly fit the orbits of the CLS systems hosting one or more companions with $M\sin i \leq80~\Mjup$. We fit three-dimensional orbits, including inclination $i$ and longitude of the ascending node $\Omega$, with the aim of constraining companion true mass $M_c$ rather than $M_c \sin i$. We used the RV measurements provided in \cite{Rosenthal2021} for all systems, and we incorporated HGCA astrometric accelerations for the 53 systems in which the acceleration was significant ($\Delta\mu / \sigma(\Delta\mu) \geq 3$). For the remaining 75 systems, we supplied no astrometric data and fit only the RVs, allowing \texttt{Orvara} to marginalize over the uncertainty in inclination.

\section{Eccentricity Analysis}\label{sec:analysis}

\subsection{Survey completeness}\label{subsec:completeness}

The primary goal of CLS was to measure planet occurrence rates, a task which demands careful appraisal of search completeness. As such, \citet{Rosenthal2021} measured CLS system-by-system sensitivity to planetary, sub-stellar, and stellar companions through a suite of injection-and-recovery tests. These tests were performed by randomly drawing a set of orbital parameters ($P$, $t_p$, $e$, $\omega$, $K$), forward-modeling the RV signature that a companion with those parameters would produce, injecting that signature into a system's actual RV time series to capture system-specific RV noise and sampling inhomogeneities, and attempting to recover the signal with a signal detection pipeline. \citet{Rosenthal2021} repeated this procedure 1000 times for each system, and estimated sensitivity as the fraction of successful recoveries in a given region of $M_p$--$a$ space. This procedure facilitated high-confidence measurement of giant planet occurrence near the ice line \citep{Fulton2021} and the conditional occurrence rate of these giant planets in the presence of close-in small planets \citep[$M_p = 2-30 \Me, a < 1 \text{AU}$;][]{Rosenthal2022}.

CLS injection-and-recovery tests assumed a beta distribution for eccentricity and marginalized over $e$ in their completeness maps. This procedure was adequate to determine occurrence rates, but may still introduce subtle biases to the inferred eccentricity distribution of the population. So, to more thoroughly explore the role of eccentricity in survey completeness, we retrieved the original injection-and-recovery results calculated by \citet{Rosenthal2021} and divided them into bins according to eccentricity ($e = 0.0 - 0.9, \Delta e = 0.1$). For each discrete value of $e$, we computed a completeness map following established procedures that treat occurrence as a censored Poisson point process \citep[][Van Zandt et al., in review]{ForemanMackey2014, Fulton2021, VanZandtPetigura2024}. Finally, we interpolated these completeness maps to determine detection efficiency $\hat{d}_{nk}$ for each of the $\{M_c, a, e\}_{nk}$ samples returned by our joint RV-astrometry fits.

\begin{figure}
    \centering
    \includegraphics[width=0.95\linewidth]{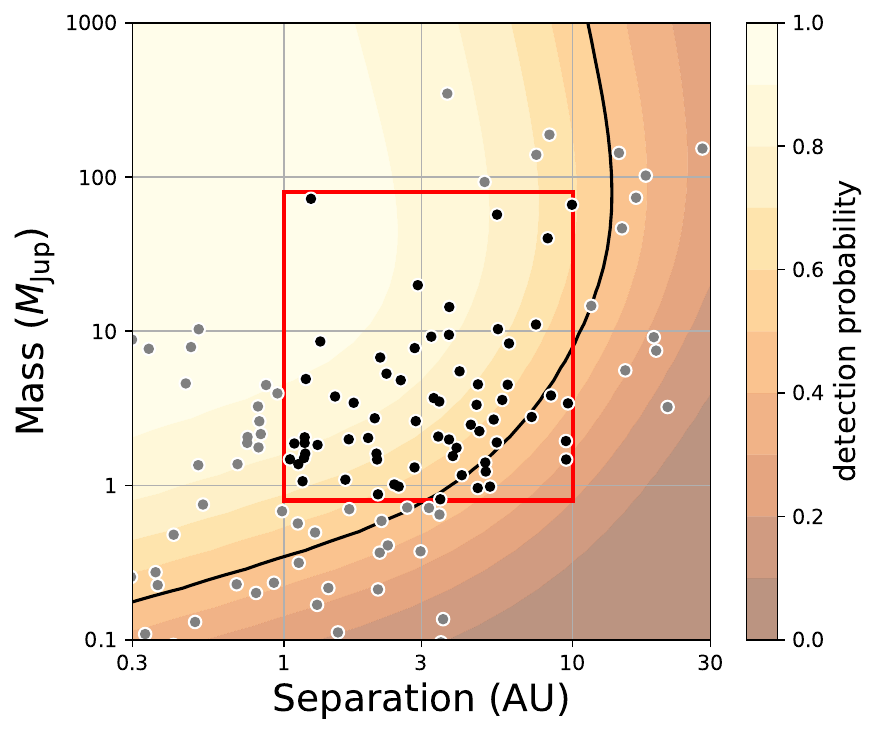}
    \caption{Masses and orbital separations of the sub-stellar companions detected by CLS. Black points indicate objects included in our study, whereas gray points are shown for context but not included in our analysis. The colored contours indicate the detection probability for a sub-stellar companion with a given mass and orbital separation and a nominal eccentricity $e=0.2$. The black contour shows the 50\% sensitivity line. The red rectangle indicates our chosen parameter space for this study, where completeness is high.}
    \label{fig:mass-separation}
\end{figure}

\subsection{Hierarchical population inference}\label{subsec:eccentricity}

\begin{figure*}
    \centering
    \includegraphics[width=0.95\linewidth]{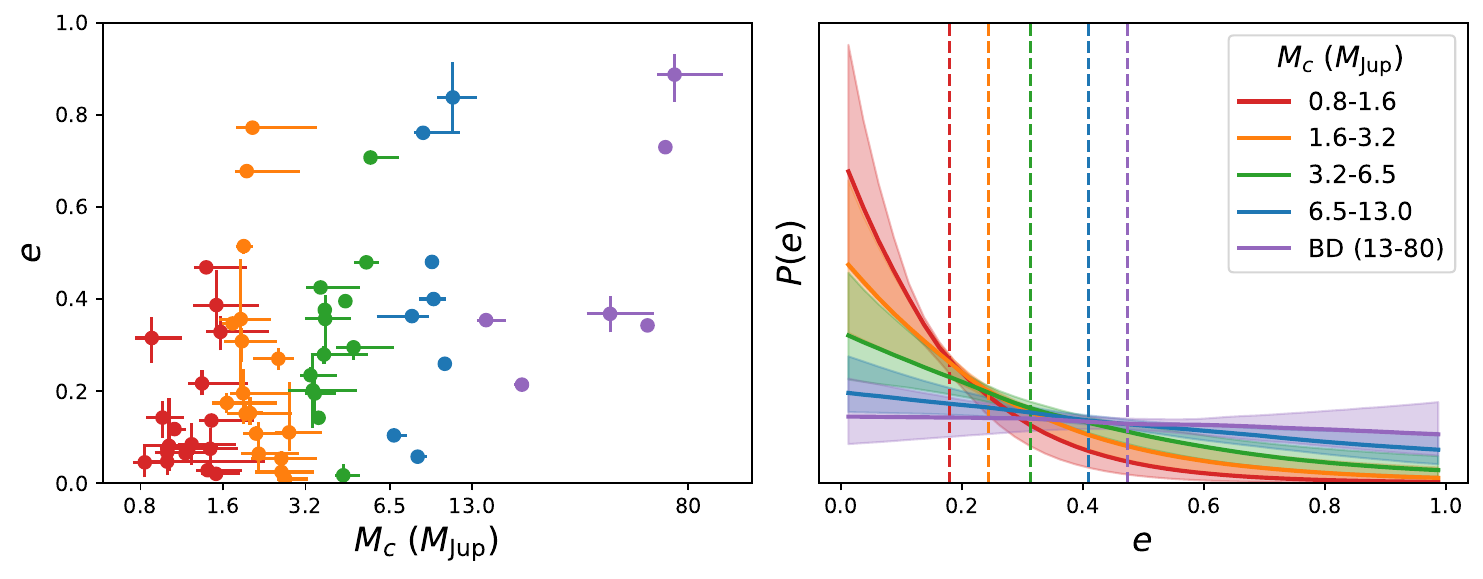}
    \caption{The eccentricities of sub-stellar objects with star-companion orbital separations $a = 1-10 \text{AU}$ and true masses $M_c = 0.8-80 \Mjup$. \textit{Left panel}: Mass vs. eccentricity. Colors correspond to our nominal size bins. There is a clear trend of rising eccentricity with rising mass, with objects below $M=1.6 \Mjup$ preferring $e \lesssim 0.5$ and objects above $M = 13 \Mjup$ preferring $e \gtrsim 0.2$. \textit{Right panel}: Sub-population eccentricity distributions inferred using a hierarchical Bayesian model. Solid lines indicate the median posterior distribution, shaded regions indicate $1\sigma$ confidence intervals, and dashed lines indicated the mean for each subpopulation. There is a smooth transition in the mean eccentricity $\e$ and the overall shape of the distribution from the lowest mass bin (peaked at $e=0$ and monotonically declining toward zero at $e=1$) to the highest pass bin (nearly uniform over $e$).}
    \label{fig:eccentricity}
\end{figure*}

By eye, there is a clear trend of rising eccentricity with increasing companion mass (Figure \ref{fig:eccentricity}). To precisely quantify this trend, we adopt the approximate hierarchical Bayesian modeling framework used by \citet{Gilbert2025} to infer the eccentricity distribution of transiting planets. Our procedure builds upon the formalism originally described by \citet{Hogg2010} and further developed by \citet{ForemanMackey2014}, \citet{VanEylen2019}, and \citet{Masuda2022}. A summary of our procedure is as follows.

Our joint RV--astrometry fits for $N$ planets each produced a set of $K$ posterior eccentricity samples $e_{nk}$. For the majority of targets, eccentricity values were tightly constrained, with $\sigma(e) \lesssim 0.1$. The combined likelihood for the population eccentricity distribution $f(e)$ is
\begin{equation}
    \mathcal{L}(e;\alpha) = \prod_{n=1}^N \frac{1}{K} \sum_{k=1}^K \frac{f(e_{nk}; \alpha)}{p_0(e_{nk})} \hat{d}(e_{nk}),
\end{equation}
where $p_0(e)$ is the uninformative ``interim'' prior on $e$ applied during model fitting and $f(e;\alpha)$ is the informative ``updated'' population distribution we wish to infer. For our purposes, $p_0(e) = 1$, since $e$ was sampled assuming a uniform prior. The quantity $\alpha$ is the vector of hyper-parameters describing the updated population distribution, and $\hat{d}(e_{nk})$ weights the samples to account for detection completeness (see \S\ref{subsec:completeness} and Van Zandt et al., in review). We performed the analysis both with and without including this completeness correction and found it made virtually no difference in the inferred eccentricity distributions. Such insensitivity to completeness corrections is unsurprising, as we chose this region intentionally for its high completeness.

Following \citet{Gilbert2025}, we adopted a flexible non-parametric model for $f(e;\alpha)$ which describes the eccentricity distribution as a regularized histogram. The advantage of using a regularized histogram is that it makes no assumptions about the shape or functional form of the distribution other than requiring that the distribution be smooth and continuous.

For a histogram with $M$ bins, the function is defined as
\begin{equation}
f(e;\alpha) = 
\sum_{m=1}^M \exp(\alpha_m) 
\gamma(e;a_m,b_m)
\end{equation}
where the hyper-parameters $\{\alpha_m\}$ are the logarithmic bin heights, ${\{a_m,b_m\}}$ define the fixed locations of the histogram bin edges such that each of the bins has width $\Delta e_m = b_m - a_m$, and $\gamma(e)$ is a correction factor
\begin{equation}
    \gamma(e;a,b) = 
    \begin{cases}
        \frac{1}{b-a},& \text{if } a \leq e < b \\
        0,              & \text{otherwise}
    \end{cases}
\end{equation}
which accounts for the finite bin widths.

To ensure proper normalization of the probability density function $f(e)$ we transformed $\alpha_m$ via
\begin{equation}
    \alpha_m' = \alpha_m - \ln\left({\sum_{m=1}^M \exp(\alpha_m)\Delta e_m}\right)
\end{equation}
so that $\sum_m{\exp(\alpha'_m)\Delta e_m} = 1$. We used 40 bins, so each bin had a width $\Delta e_m = 0.025$.

We enforced smoothness in the histogram by applying a Gaussian Process (GP) prior on $\alpha$ using a Mat\'ern-3/2 kernel. The covariance function of this kernel is
\begin{equation}\label{eq:matern32}
    \kappa(e) = s^2\left(1 + \frac{\sqrt{3}e}{\ell}\right)\exp\left(-\frac{\sqrt{3}e}{\ell}\right)
\end{equation}
where $s$ sets the scale and $\ell$ sets the correlation length. In practice, we achieved good results by applying a modestly informative log-normal hyper-priors on the GP hyperparameters $\ln(s)\sim \mathcal{N}(3,1)$ and $\ln(\ell)\sim \mathcal{N}(0,1)$. These hyper-priors prevented the sampler from wandering too far from the maximum likelihood region of parameter space and enforced a certain degree of smoothness in the distribution. Latent bin heights $\alpha$ were drawn from a standard normal distribution and scaled according to the GP prior.

To determine the relationship between $M_c$ and $f(e)$, we split the sub-stellar object population into five approximately log-uniform mass bins, with bin edges at $M_c/\Mjup = [0.8, 1.6, 3.2, 6.5, 13, 80]$; the larger bin size for classical brown dwarfs was chosen to accommodate the small sample size for these objects. To maximize information content, Gaussian process regularization hyperparameters $\{s, \ell\}$ were shared between the five groups, which is equivalent to assuming that the various sub-populations have eccentricity distributions with roughly comparable smoothness. Other than this mild assumption, no further relationship between the subpopulations was imposed. We sampled from the posterior using \texttt{PyMC3} and a gradient-based Hamiltonian Monte Carlo implementation of the No U-Turn Sampling (NUTS) algorithm \citep{HoffmanGelman2011, BetancourtGirolami2013, pymc3}. Our fitting algorithm produced posterior samples for the intrinsic distribution $f(e)$ for each of the five sub-populations.

Eccentricities for individual objects and sub-populations binned by sub-stellar object mass are shown in Figure \ref{fig:eccentricity}. From the raw data alone (left panel), we observed a clear trend of rising eccentricity with rising mass, with objects below $M=1.6 \Mjup$ preferring $e \lesssim 0.5$ and objects above $M = 13 \Mjup$ preferring $e \gtrsim 0.2$. Our hierarchical inference (right panel) further revealed that there is a smooth transition in the mean eccentricity $\e$ and the overall shape of the distribution from the lowest mass bin to the highest mass bin). Furthermore, the relationship between $M_c$ and $e$ appears to be monotonic. The distribution the lowest mass bin is peaked at $e=0$ and falls quasi-exponentially toward zero at $e=1$, transitioning gradually toward a nearly uniform distribution for the nominal deuterium-burning brown dwarfs.

\section{Discussion}\label{sec:discussion}

Our hierarchical inference of population eccentricity distributions suggests that the transition from low-mass, low-eccentricity objects to high-mass, varied-eccentricity objects is gradual between $1-10 \Mjup$. This positive correlation between $M_c$ and $\e$ is expected from theoretical models of strong gravitational scattering, and does not necessarily indicate that the higher mass objects formed in a qualitatively distinct manner from the lower mass objects. \citep{Chatterjee2008, FordRasio2008, JuricTremaine2008, Ida2013-eccentricity}. Moreover, no apparent upper limit on eccentricities that may be formed by planet-planet scattering, although proto-planetary architectures which produce $e \gtrsim 0.8$ via gravitational scattering are rare \citep{Carrera2019}.

\begin{figure*}
    \centering
    \includegraphics[width=0.85\linewidth]{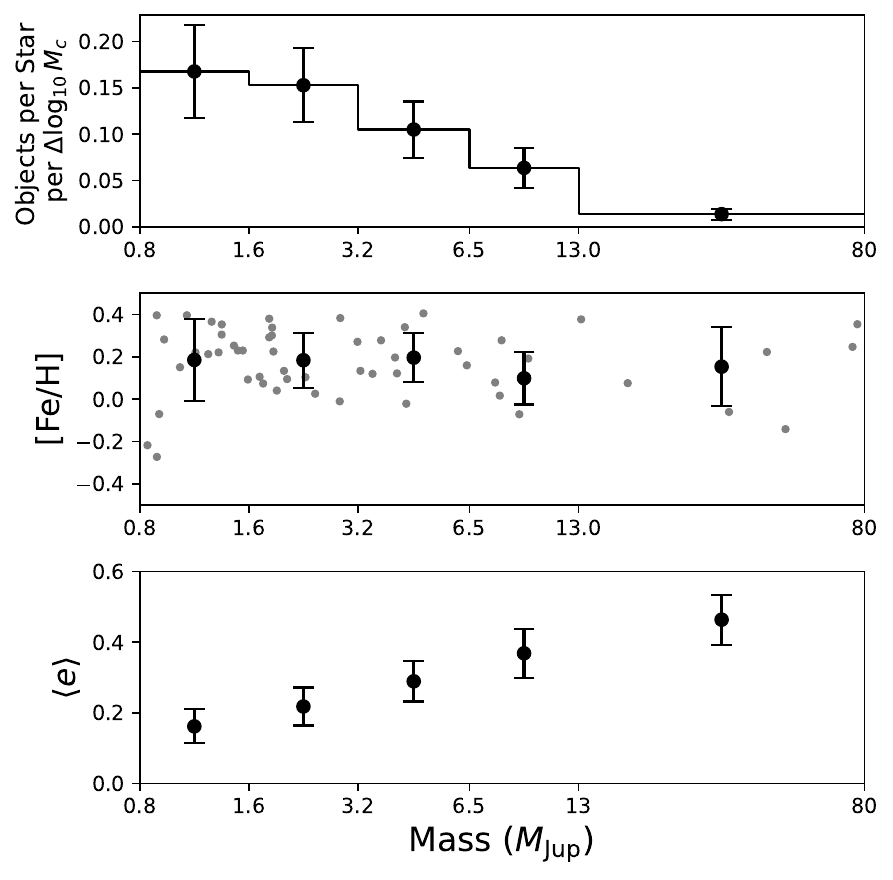}
    \caption{Demographic trends in occurrence rate density, host star metallicity, and orbital eccentricity for sub-stellar objects between $0.8-80\Mjup$. All three demographic trends display a smooth variation with companion mass. \textit{Top panel}: Occurrence rate per logarithmic mass bin falls steadily as $M_c$ increases. \textit{Middle panel}: The relationship between host star metallicity and companion mass is weak-to-nonexistent, but may be weakly anti-correlated. \textit{Bottom panel}: Mean eccentricity $\e$ rises steadily as $M_c$ increases.}
    \label{fig:demographics}
\end{figure*}

To place our new eccentricity measurements in context, we consider each sub-population's mean eccentricity $\e$ alongside host star metallicity [Fe/H] and companion occurrence rate density (see Figure \ref{fig:demographics}). We found that whereas occurrence rate density steadily decreases with increasing $M_c$ (Van Zandt et al., in review), [Fe/H] and $M_c$ exhibit no strong relationship for our sample. 

The observed anti-correlation between occurrence rate density and $M_c$ is consistent with previous analyses which found a paucity of brown dwarfs inside $a \lesssim 3$ AU \citep[i.e. the so-called ``Brown Dwarf desert'';][]{GretherLineweaver2006}. The lack of a correlation between [Fe/H] and $M_c$, however, stands in tension with previous studies which have claimed to identify a transition in host star metallicity and $M_c$ between $4-10\Mjup$ \citep{Santos2017, Schlaufman2018, Adibekyan2019}. Nevertheless, we caution against over-interpreting this tension, as the comparison is not strictly one-to-one. For example, \citet{Schlaufman2018} placed the metallicity-derived dividing line at $M_c \approx 4\Mjup$, grouping objects between $0.2 < M_c/\Mjup < 4$ vs $4 < M_c/\Mjup < 300$, meaning their analysis included smaller planets in the first group and both large brown dwarfs and very low mass stars in the second group. In addition, these previous analyses focused on samples of close-in objects, whereas our sample extends out to 10 AU, near the orbit of Saturn. A companion study to this work, Giacalone et al. (in review) places the transition at $M_c = 25^{+9}_{-7} \Mjup$ for the CLS sample, with a broad range of permissible values ($95\%$ confidence interval [14,48] $\Mjup$). So, our non-detection of a threshold mass below $13 \Mjup$ where host star [Fe/H] abruptly changes is not an outright contraction. 

The lack of a sharp transition in occurrence, host star metallicity, and orbital eccentricity between giant planets and brown dwarfs is qualitatively different than the sharp transition in these quantities between small planets ($R_p < 3.5 \Re$) and large planets ($R_p > 3.5\Re$) found by \citet{Gilbert2025}. Whereas giant planet formation appears to be a threshold process --- i.e. an ``all-or-nothing'' event predicated on whether a protoplanetary core reaches the requisite mass for runaway H/He envelope accretion \citep{Pollack1996} --- there appears to be no such threshold governing whether or if a super-giant planet turns into a brown dwarf. The simplest explanation for our non-detection of a sharp threshold is that both bottom-up and top-down formation mechanisms produce objects in the range $1-10 \Mjup$. Exactly how large of an object can be formed by core accretion or how small of an object can be formed by disk instability or cloud fragmentation remains to be determined. Nevertheless, theoretical models suggest that sub-stellar companions with masses $M_c \sim 10 \Mjup$ can be formed under rare but feasible circumstances via both bottom-up \citep{Tanaka2020} and top-down channels \citep{Stamatellos2007, DodsonRobinson2009, Boss2011, VorobyovElbakyan2018}. So, we may reasonably assume that as the mass of an object increases, the likelihood that it formed via core accretion drops and the likelihood that it formed by gravitational instability rises.

\section{Conclusion}\label{sec:conclusion}

We analyzed the eccentricities of 70 sub-stellar objects with masses between $0.8 - 80 \Mjup$ using a hierarchical Bayesian framework. These objects were uniformly detected and vetted by the California Legacy Survey (CLS), enabling completeness corrections and unbiased population inference. Our main conclusions are as follows.

\begin{itemize}
    \item Doppler-detected Jovian-mass objects between 1--10 AU have $\e \approx 0.2$ whereas brown dwarfs above the deuterium burning limit have $\e \approx 0.5$.
    
    \item The transition in $\e$ and $f(e)$ from low-mass to high-mass objects is gradual over this mass range.
    
    \item Changes in host star metallicity and companion occurrence rates are also gradual in this regime. These trends imply that bottom-up formation mechanisms (i.e. core accretion) and top-down mechanisms (i.e. gravitational instability) produce sub-stellar objects that overlap in mass.
\end{itemize}

Our results are consistent with numerous previous analyses which demonstrated that giant planets tend to have low eccentricities with a relatively narrow spread in values \citep{Kipping2013, Bowler2020, Gilbert2025}, whereas brown dwarfs tend to have a more nearly uniform distribution of eccentricities \citep{MaGe2014, Bowler2020}. By focusing on a relatively narrow region of parameter space ($a = 1-10$ AU, $M_c = 0.8-80\Mjup$), we demonstrated that the transition between the giant planet regime and the brown dwarf regime is gradual, with slow-and-steady changes not only in eccentricity, but also in occurrence rate and metallicity (see Van Zandt et al., in review and Giacalone et al, in review for further discussion).

This project endeavored to identify a dynamical signature that could be used to distinguish super-giant planets from small brown dwarfs. 

Unfortunately, it appears that eccentricity alone does not provide sufficient information. Perhaps a clean dividing line between formation channels does exist, but we have not yet found it either because we do not have enough objects or because we have not yet examined the right combination of parameters. Recall, for example, how much cleaner the ``radius gap'' separating super-Earths and sub-Neptunes appears in the 2D space of period and radius rather than the 1D space of radius alone \citep{Fulton2017, FultonPetigura2018}. For giant planets and brown dwarfs, perhaps one could fold together other quantities such as star-companion spin-orbit angle, companion spin rate, companion atmospheric chemistry or other quantities alongside mass and eccentricity to find a clean separation between top-down vs bottom-up channels.

\clearpage
\section*{Acknowledgments}

We thank the anonymous referee for helpful comments which improved the quality of the manuscript. We are grateful for insightful conversations with Erik Ford and Ruth Murray-Clay which guided the interpretation of results.

Funding for this work was provided by a University of California, Los Angeles set-up award to E.A.P. and by the Heising-Simons Foundation Award 2022-3833. J.V.Z. was supported by NASA FINESST Fellowship 80NSSC22K1606.

This research has made use of the NASA Exoplanet Archive, which is operated by the California Institute of Technology, under contract with the National Aeronautics and Space Administration under the Exoplanet Exploration Program. This research has made use of the Astrophysics Data System, funded by NASA under Cooperative Agreement 80NSSC25M7105.

Some of the data presented herein were obtained at Keck Observatory, which is a private 501(c)3 non-profit organization operated as a scientific partnership among the California Institute of Technology, the University of California, and the National Aeronautics and Space Administration. The Observatory was made possible by the generous financial support of the W. M. Keck Foundation.

The authors wish to recognize and acknowledge the very significant cultural role and reverence that the summit of Maunakea has always had within the Native Hawaiian community. We are most fortunate to have the opportunity to conduct observations from this mountain.

\software{\texttt{astropy} \citep{astropy:2018},
          \texttt{numpy} \citep{numpy:2020}, 
          \texttt{scipy} \citep{scipy:2020}, 
          \texttt{pymc3} \citep{pymc3}
          }

\bibliography{main}

@ARTICLE{Adibekyan2019,
       author = {{Adibekyan}, Vardan},
        title = "{Heavy Metal Rules. I. Exoplanet Incidence and Metallicity}",
      journal = {Geosciences},
         year = 2019,
        month = feb,
       volume = {9},
       number = {3},
          eid = {105},
        pages = {105},
          doi = {10.3390/geosciences9030105},
archivePrefix = {arXiv},
       eprint = {1902.04493},
 primaryClass = {astro-ph.EP},
       adsurl = {https://ui.adsabs.harvard.edu/abs/2019Geosc...9..105A}
}

@ARTICLE{astropy:2018,
       author = {{Astropy Collaboration} and {Price-Whelan}, A.~M. and
         {Sip{\H{o}}cz}, B.~M. and {G{\"u}nther}, H.~M. and {Lim}, P.~L. and
         {Crawford}, S.~M. and {Conseil}, S. and {Shupe}, D.~L. and
         {Craig}, M.~W. and {Dencheva}, N. and {Ginsburg}, A. and {Vand
        erPlas}, J.~T. and {Bradley}, L.~D. and {P{\'e}rez-Su{\'a}rez}, D. and
         {de Val-Borro}, M. and {Aldcroft}, T.~L. and {Cruz}, K.~L. and
         {Robitaille}, T.~P. and {Tollerud}, E.~J. and {Ardelean}, C. and
         {Babej}, T. and {Bach}, Y.~P. and {Bachetti}, M. and {Bakanov}, A.~V. and
         {Bamford}, S.~P. and {Barentsen}, G. and {Barmby}, P. and
         {Baumbach}, A. and {Berry}, K.~L. and {Biscani}, F. and {Boquien}, M. and
         {Bostroem}, K.~A. and {Bouma}, L.~G. and {Brammer}, G.~B. and
         {Bray}, E.~M. and {Breytenbach}, H. and {Buddelmeijer}, H. and
         {Burke}, D.~J. and {Calderone}, G. and {Cano Rodr{\'\i}guez}, J.~L. and
         {Cara}, M. and {Cardoso}, J.~V.~M. and {Cheedella}, S. and {Copin}, Y. and
         {Corrales}, L. and {Crichton}, D. and {D'Avella}, D. and {Deil}, C. and
         {Depagne}, {\'E}. and {Dietrich}, J.~P. and {Donath}, A. and
         {Droettboom}, M. and {Earl}, N. and {Erben}, T. and {Fabbro}, S. and
         {Ferreira}, L.~A. and {Finethy}, T. and {Fox}, R.~T. and
         {Garrison}, L.~H. and {Gibbons}, S.~L.~J. and {Goldstein}, D.~A. and
         {Gommers}, R. and {Greco}, J.~P. and {Greenfield}, P. and
         {Groener}, A.~M. and {Grollier}, F. and {Hagen}, A. and {Hirst}, P. and
         {Homeier}, D. and {Horton}, A.~J. and {Hosseinzadeh}, G. and {Hu}, L. and
         {Hunkeler}, J.~S. and {Ivezi{\'c}}, {\v{Z}}. and {Jain}, A. and
         {Jenness}, T. and {Kanarek}, G. and {Kendrew}, S. and {Kern}, N.~S. and
         {Kerzendorf}, W.~E. and {Khvalko}, A. and {King}, J. and {Kirkby}, D. and
         {Kulkarni}, A.~M. and {Kumar}, A. and {Lee}, A. and {Lenz}, D. and
         {Littlefair}, S.~P. and {Ma}, Z. and {Macleod}, D.~M. and
         {Mastropietro}, M. and {McCully}, C. and {Montagnac}, S. and
         {Morris}, B.~M. and {Mueller}, M. and {Mumford}, S.~J. and {Muna}, D. and
         {Murphy}, N.~A. and {Nelson}, S. and {Nguyen}, G.~H. and
         {Ninan}, J.~P. and {N{\"o}the}, M. and {Ogaz}, S. and {Oh}, S. and
         {Parejko}, J.~K. and {Parley}, N. and {Pascual}, S. and {Patil}, R. and
         {Patil}, A.~A. and {Plunkett}, A.~L. and {Prochaska}, J.~X. and
         {Rastogi}, T. and {Reddy Janga}, V. and {Sabater}, J. and
         {Sakurikar}, P. and {Seifert}, M. and {Sherbert}, L.~E. and
         {Sherwood-Taylor}, H. and {Shih}, A.~Y. and {Sick}, J. and
         {Silbiger}, M.~T. and {Singanamalla}, S. and {Singer}, L.~P. and
         {Sladen}, P.~H. and {Sooley}, K.~A. and {Sornarajah}, S. and
         {Streicher}, O. and {Teuben}, P. and {Thomas}, S.~W. and
         {Tremblay}, G.~R. and {Turner}, J.~E.~H. and {Terr{\'o}n}, V. and
         {van Kerkwijk}, M.~H. and {de la Vega}, A. and {Watkins}, L.~L. and
         {Weaver}, B.~A. and {Whitmore}, J.~B. and {Woillez}, J. and
         {Zabalza}, V. and {Astropy Contributors}},
        title = "{The Astropy Project: Building an Open-science Project and Status of the v2.0 Core Package}",
      journal = {\aj},
         year = 2018,
        month = sep,
       volume = {156},
       number = {3},
          eid = {123},
        pages = {123},
          doi = {10.3847/1538-3881/aabc4f},
archivePrefix = {arXiv},
       eprint = {1801.02634},
 primaryClass = {astro-ph.IM},
       adsurl = {https://ui.adsabs.harvard.edu/abs/2018AJ....156..123A}
}

@ARTICLE{BetancourtGirolami2013,
       author = {{Betancourt}, M.~J. and {Girolami}, Mark},
        title = "{Hamiltonian Monte Carlo for Hierarchical Models}",
      journal = {arXiv e-prints},
         year = 2013,
        month = dec,
          eid = {arXiv:1312.0906},
        pages = {arXiv:1312.0906},
          doi = {10.48550/arXiv.1312.0906},
archivePrefix = {arXiv},
       eprint = {1312.0906},
 primaryClass = {stat.ME},
       adsurl = {https://ui.adsabs.harvard.edu/abs/2013arXiv1312.0906B}
}

@ARTICLE{Boss1997,
       author = {{Boss}, A.~P.},
        title = "{Giant planet formation by gravitational instability.}",
      journal = {Science},
         year = 1997,
        month = jan,
       volume = {276},
        pages = {1836-1839},
          doi = {10.1126/science.276.5320.1836},
       adsurl = {https://ui.adsabs.harvard.edu/abs/1997Sci...276.1836B}
}

@ARTICLE{Boss2011,
       author = {{Boss}, Alan P.},
        title = "{Formation of Giant Planets by Disk Instability on Wide Orbits Around Protostars with Varied Masses}",
      journal = {\apj},
         year = 2011,
        month = apr,
       volume = {731},
       number = {1},
          eid = {74},
        pages = {74},
          doi = {10.1088/0004-637X/731/1/74},
archivePrefix = {arXiv},
       eprint = {1102.4555},
 primaryClass = {astro-ph.EP},
       adsurl = {https://ui.adsabs.harvard.edu/abs/2011ApJ...731...74B}
}

@ARTICLE{Bowler2020,
       author = {{Bowler}, Brendan P. and {Blunt}, Sarah C. and {Nielsen}, Eric L.},
        title = "{Population-level Eccentricity Distributions of Imaged Exoplanets and Brown Dwarf Companions: Dynamical Evidence for Distinct Formation Channels}",
      journal = {\aj},
         year = 2020,
        month = feb,
       volume = {159},
       number = {2},
          eid = {63},
        pages = {63},
          doi = {10.3847/1538-3881/ab5b11},
archivePrefix = {arXiv},
       eprint = {1911.10569},
 primaryClass = {astro-ph.EP},
       adsurl = {https://ui.adsabs.harvard.edu/abs/2020AJ....159...63B}
}

@ARTICLE{Buchhave2012,
       author = {{Buchhave}, Lars A. and {Latham}, David W. and {Johansen}, Anders and {Bizzarro}, Martin and {Torres}, Guillermo and {Rowe}, Jason F. and {Batalha}, Natalie M. and {Borucki}, William J. and {Brugamyer}, Erik and {Caldwell}, Caroline and {Bryson}, Stephen T. and {Ciardi}, David R. and {Cochran}, William D. and {Endl}, Michael and {Esquerdo}, Gilbert A. and {Ford}, Eric B. and {Geary}, John C. and {Gilliland}, Ronald L. and {Hansen}, Terese and {Isaacson}, Howard and {Laird}, John B. and {Lucas}, Philip W. and {Marcy}, Geoffrey W. and {Morse}, Jon A. and {Robertson}, Paul and {Shporer}, Avi and {Stefanik}, Robert P. and {Still}, Martin and {Quinn}, Samuel N.},
        title = "{An abundance of small exoplanets around stars with a wide range of metallicities}",
      journal = {\nat},
         year = 2012,
        month = jun,
       volume = {486},
       number = {7403},
        pages = {375-377},
          doi = {10.1038/nature11121},
       adsurl = {https://ui.adsabs.harvard.edu/abs/2012Natur.486..375B}
}

@ARTICLE{Orvara2021,
       author = {{Brandt}, Timothy D. and {Dupuy}, Trent J. and {Li}, Yiting and {Brandt}, G. Mirek and {Zeng}, Yunlin and {Michalik}, Daniel and {Bardalez Gagliuffi}, Daniella C. and {Raposo-Pulido}, Virginia},
        title = "{orvara: An Efficient Code to Fit Orbits Using Radial Velocity, Absolute, and/or Relative Astrometry}",
      journal = {\aj},
         year = 2021,
        month = nov,
       volume = {162},
       number = {5},
          eid = {186},
        pages = {186},
          doi = {10.3847/1538-3881/ac042e},
archivePrefix = {arXiv},
       eprint = {2105.11671},
 primaryClass = {astro-ph.IM},
       adsurl = {https://ui.adsabs.harvard.edu/abs/2021AJ....162..186B}
}

@ARTICLE{Brandt2021,
       author = {{Brandt}, Timothy D.},
        title = "{The Hipparcos-Gaia Catalog of Accelerations: Gaia EDR3 Edition}",
      journal = {\apjs},
         year = 2021,
        month = jun,
       volume = {254},
       number = {2},
          eid = {42},
        pages = {42},
          doi = {10.3847/1538-4365/abf93c},
archivePrefix = {arXiv},
       eprint = {2105.11662},
 primaryClass = {astro-ph.GA},
       adsurl = {https://ui.adsabs.harvard.edu/abs/2021ApJS..254...42B}
}

@ARTICLE{Carrera2019,
       author = {{Carrera}, Daniel and {Raymond}, Sean N. and {Davies}, Melvyn B.},
        title = "{Planet-planet scattering as the source of the highest eccentricity exoplanets}",
      journal = {\aap},
         year = 2019,
        month = sep,
       volume = {629},
          eid = {L7},
        pages = {L7},
          doi = {10.1051/0004-6361/201935744},
archivePrefix = {arXiv},
       eprint = {1903.02564},
 primaryClass = {astro-ph.EP},
       adsurl = {https://ui.adsabs.harvard.edu/abs/2019A&A...629L...7C}
}

@ARTICLE{Chatterjee2008,
       author = {{Chatterjee}, Sourav and {Ford}, Eric B. and {Matsumura}, Soko and {Rasio}, Frederic A.},
        title = "{Dynamical Outcomes of Planet-Planet Scattering}",
      journal = {\apj},
         year = 2008,
        month = oct,
       volume = {686},
       number = {1},
        pages = {580-602},
          doi = {10.1086/590227},
archivePrefix = {arXiv},
       eprint = {astro-ph/0703166},
 primaryClass = {astro-ph},
       adsurl = {https://ui.adsabs.harvard.edu/abs/2008ApJ...686..580C}
}

@ARTICLE{DoO2023,
       author = {{Do {\'O}}, Clarissa R. and {O'Neil}, Kelly K. and {Konopacky}, Quinn M. and {Do}, Tuan and {Martinez}, Gregory D. and {Ruffio}, Jean-Baptiste and {Ghez}, Andrea M.},
        title = "{The Orbital Eccentricities of Directly Imaged Companions Using Observable-based Priors: Implications for Population-level Distributions}",
      journal = {\aj},
         year = 2023,
        month = aug,
       volume = {166},
       number = {2},
          eid = {48},
        pages = {48},
          doi = {10.3847/1538-3881/acdc9a},
archivePrefix = {arXiv},
       eprint = {2306.04080},
 primaryClass = {astro-ph.EP},
       adsurl = {https://ui.adsabs.harvard.edu/abs/2023AJ....166...48D}
}

@INPROCEEDINGS{Chabrier2014,
       author = {{Chabrier}, G. and {Johansen}, A. and {Janson}, M. and {Rafikov}, R.},
        title = "{Giant Planet and Brown Dwarf Formation}",
    booktitle = {Protostars and Planets VI},
         year = 2014,
       editor = {{Beuther}, Henrik and {Klessen}, Ralf S. and {Dullemond}, Cornelis P. and {Henning}, Thomas},
        month = jan,
        pages = {619-642},
          doi = {10.2458/azu_uapress_9780816531240-ch027},
archivePrefix = {arXiv},
       eprint = {1401.7559},
 primaryClass = {astro-ph.SR},
       adsurl = {https://ui.adsabs.harvard.edu/abs/2014prpl.conf..619C}
}

@ARTICLE{DodsonRobinson2009,
       author = {{Dodson-Robinson}, Sarah E. and {Veras}, Dimitri and {Ford}, Eric B. and {Beichman}, C.~A.},
        title = "{The Formation Mechanism of Gas Giants on Wide Orbits}",
      journal = {\apj},
         year = 2009,
        month = dec,
       volume = {707},
       number = {1},
        pages = {79-88},
          doi = {10.1088/0004-637X/707/1/79},
archivePrefix = {arXiv},
       eprint = {0909.2662},
 primaryClass = {astro-ph.EP},
       adsurl = {https://ui.adsabs.harvard.edu/abs/2009ApJ...707...79D}
}

@ARTICLE{FischerValenti2005,
       author = {{Fischer}, Debra A. and {Valenti}, Jeff},
        title = "{The Planet-Metallicity Correlation}",
      journal = {\apj},
         year = 2005,
        month = apr,
       volume = {622},
       number = {2},
        pages = {1102-1117},
          doi = {10.1086/428383},
       adsurl = {https://ui.adsabs.harvard.edu/abs/2005ApJ...622.1102F}
}

@ARTICLE{FordRasio2008,
       author = {{Ford}, Eric B. and {Rasio}, Frederic A.},
        title = "{Origins of Eccentric Extrasolar Planets: Testing the Planet-Planet Scattering Model}",
      journal = {\apj},
         year = 2008,
        month = oct,
       volume = {686},
       number = {1},
        pages = {621-636},
          doi = {10.1086/590926},
archivePrefix = {arXiv},
       eprint = {astro-ph/0703163},
 primaryClass = {astro-ph},
       adsurl = {https://ui.adsabs.harvard.edu/abs/2008ApJ...686..621F}
}

@ARTICLE{ForemanMackey2014,
       author = {{Foreman-Mackey}, Daniel and {Hogg}, David W. and {Morton}, Timothy D.},
        title = "{Exoplanet Population Inference and the Abundance of Earth Analogs from Noisy, Incomplete Catalogs}",
      journal = {ApJ},
         year = 2014,
        month = nov,
       volume = {795},
       number = {1},
          eid = {64},
        pages = {64},
          doi = {10.1088/0004-637X/795/1/64},
archivePrefix = {arXiv},
       eprint = {1406.3020},
 primaryClass = {astro-ph.EP},
       adsurl = {https://ui.adsabs.harvard.edu/abs/2014ApJ...795...64F}
}

@ARTICLE{ForganRice2011,
       author = {{Forgan}, Duncan and {Rice}, Ken},
        title = "{The Jeans mass as a fundamental measure of self-gravitating disc fragmentation and initial fragment mass}",
      journal = {\mnras},
         year = 2011,
        month = nov,
       volume = {417},
       number = {3},
        pages = {1928-1937},
          doi = {10.1111/j.1365-2966.2011.19380.x},
archivePrefix = {arXiv},
       eprint = {1107.0831},
 primaryClass = {astro-ph.EP},
       adsurl = {https://ui.adsabs.harvard.edu/abs/2011MNRAS.417.1928F}
}

@ARTICLE{Fulton2017,
       author = {{Fulton}, Benjamin J. and {Petigura}, Erik A. and {Howard}, Andrew W. and {Isaacson}, Howard and {Marcy}, Geoffrey W. and {Cargile}, Phillip A. and {Hebb}, Leslie and {Weiss}, Lauren M. and {Johnson}, John Asher and {Morton}, Timothy D. and {Sinukoff}, Evan and {Crossfield}, Ian J.~M. and {Hirsch}, Lea A.},
        title = "{The California-Kepler Survey. III. A Gap in the Radius Distribution of Small Planets}",
      journal = {\aj},
         year = 2017,
        month = sep,
       volume = {154},
       number = {3},
          eid = {109},
        pages = {109},
          doi = {10.3847/1538-3881/aa80eb},
archivePrefix = {arXiv},
       eprint = {1703.10375},
 primaryClass = {astro-ph.EP},
       adsurl = {https://ui.adsabs.harvard.edu/abs/2017AJ....154..109F}
}

@ARTICLE{FultonPetigura2018,
       author = {{Fulton}, Benjamin J. and {Petigura}, Erik A.},
        title = "{The California-Kepler Survey. VII. Precise Planet Radii Leveraging Gaia DR2 Reveal the Stellar Mass Dependence of the Planet Radius Gap}",
      journal = {\aj},
         year = 2018,
        month = dec,
       volume = {156},
       number = {6},
          eid = {264},
        pages = {264},
          doi = {10.3847/1538-3881/aae828},
archivePrefix = {arXiv},
       eprint = {1805.01453},
 primaryClass = {astro-ph.EP},
       adsurl = {https://ui.adsabs.harvard.edu/abs/2018AJ....156..264F}
}

@ARTICLE{Fulton2021,
       author = {{Fulton}, Benjamin J. and {Rosenthal}, Lee J. and {Hirsch}, Lea A. and {Isaacson}, Howard and {Howard}, Andrew W. and {Dedrick}, Cayla M. and {Sherstyuk}, Ilya A. and {Blunt}, Sarah C. and {Petigura}, Erik A. and {Knutson}, Heather A. and {Behmard}, Aida and {Chontos}, Ashley and {Crepp}, Justin R. and {Crossfield}, Ian J.~M. and {Dalba}, Paul A. and {Fischer}, Debra A. and {Henry}, Gregory W. and {Kane}, Stephen R. and {Kosiarek}, Molly and {Marcy}, Geoffrey W. and {Rubenzahl}, Ryan A. and {Weiss}, Lauren M. and {Wright}, Jason T.},
        title = "{California Legacy Survey. II. Occurrence of Giant Planets beyond the Ice Line}",
      journal = {\apjs},
         year = 2021,
        month = jul,
       volume = {255},
       number = {1},
          eid = {14},
        pages = {14},
          doi = {10.3847/1538-4365/abfcc1},
archivePrefix = {arXiv},
       eprint = {2105.11584},
}

@ARTICLE{GaiaEDR3,
       author = {{Lindegren}, L. and {Klioner}, S.~A. and {Hern{\'a}ndez}, J. and {Bombrun}, A. and {Ramos-Lerate}, M. and {Steidelm{\"u}ller}, H. and {Bastian}, U. and {Biermann}, M. and {de Torres}, A. and {Gerlach}, E. and {Geyer}, R. and {Hilger}, T. and {Hobbs}, D. and {Lammers}, U. and {McMillan}, P.~J. and {Stephenson}, C.~A. and {Casta{\~n}eda}, J. and {Davidson}, M. and {Fabricius}, C. and {Gracia-Abril}, G. and {Portell}, J. and {Rowell}, N. and {Teyssier}, D. and {Torra}, F. and {Bartolom{\'e}}, S. and {Clotet}, M. and {Garralda}, N. and {Gonz{\'a}lez-Vidal}, J.~J. and {Torra}, J. and {Abbas}, U. and {Altmann}, M. and {Anglada Varela}, E. and {Balaguer-N{\'u}{\~n}ez}, L. and {Balog}, Z. and {Barache}, C. and {Becciani}, U. and {Bernet}, M. and {Bertone}, S. and {Bianchi}, L. and {Bouquillon}, S. and {Brown}, A.~G.~A. and {Bucciarelli}, B. and {Busonero}, D. and {Butkevich}, A.~G. and {Buzzi}, R. and {Cancelliere}, R. and {Carlucci}, T. and {Charlot}, P. and {Cioni}, M. -R.~L. and {Crosta}, M. and {Crowley}, C. and {del Peloso}, E.~F. and {del Pozo}, E. and {Drimmel}, R. and {Esquej}, P. and {Fienga}, A. and {Fraile}, E. and {Gai}, M. and {Garcia-Reinaldos}, M. and {Guerra}, R. and {Hambly}, N.~C. and {Hauser}, M. and {Jan{\ss}en}, K. and {Jordan}, S. and {Kostrzewa-Rutkowska}, Z. and {Lattanzi}, M.~G. and {Liao}, S. and {Licata}, E. and {Lister}, T.~A. and {L{\"o}ffler}, W. and {Marchant}, J.~M. and {Masip}, A. and {Mignard}, F. and {Mints}, A. and {Molina}, D. and {Mora}, A. and {Morbidelli}, R. and {Murphy}, C.~P. and {Pagani}, C. and {Panuzzo}, P. and {Pe{\~n}alosa Esteller}, X. and {Poggio}, E. and {Re Fiorentin}, P. and {Riva}, A. and {Sagrist{\`a} Sell{\'e}s}, A. and {Sanchez Gimenez}, V. and {Sarasso}, M. and {Sciacca}, E. and {Siddiqui}, H.~I. and {Smart}, R.~L. and {Souami}, D. and {Spagna}, A. and {Steele}, I.~A. and {Taris}, F. and {Utrilla}, E. and {van Reeven}, W. and {Vecchiato}, A.},
        title = "{Gaia Early Data Release 3. The astrometric solution}",
      journal = {\aap},
         year = 2021,
        month = may,
       volume = {649},
          eid = {A2},
        pages = {A2},
          doi = {10.1051/0004-6361/202039709},
archivePrefix = {arXiv},
       eprint = {2012.03380},
 primaryClass = {astro-ph.IM},
       adsurl = {https://ui.adsabs.harvard.edu/abs/2021A&A...649A...2L}
}

@ARTICLE{Gilbert2025,
       author = {{Gilbert}, Gregory J. and {Petigura}, Erik A. and {Entrican}, Paige M.},
        title = "{Planets larger than Neptune have elevated eccentricities}",
      journal = {Proceedings of the National Academy of Science},
         year = 2025,
        month = mar,
       volume = {122},
       number = {11},
          eid = {e2405295122},
        pages = {e2405295122},
          doi = {10.1073/pnas.2405295122},
archivePrefix = {arXiv},
       eprint = {2507.07840},
 primaryClass = {astro-ph.EP},
       adsurl = {https://ui.adsabs.harvard.edu/abs/2025PNAS..12205295G}
}

@ARTICLE{GretherLineweaver2006,
       author = {{Grether}, Daniel and {Lineweaver}, Charles H.},
        title = "{How Dry is the Brown Dwarf Desert? Quantifying the Relative Number of Planets, Brown Dwarfs, and Stellar Companions around Nearby Sun-like Stars}",
      journal = {\apj},
         year = 2006,
        month = apr,
       volume = {640},
       number = {2},
        pages = {1051-1062},
          doi = {10.1086/500161},
archivePrefix = {arXiv},
       eprint = {astro-ph/0412356},
 primaryClass = {astro-ph},
       adsurl = {https://ui.adsabs.harvard.edu/abs/2006ApJ...640.1051G}
}

@PROCEEDINGS{HipparcosCatalog,
        title = "{The HIPPARCOS and TYCHO catalogues. Astrometric and photometric star catalogues derived from the ESA HIPPARCOS Space Astrometry Mission}",
    booktitle = {ESA Special Publication},
         year = 1997,
       series = {ESA Special Publication},
       volume = {1200},
        month = jan,
       adsurl = {https://ui.adsabs.harvard.edu/abs/1997ESASP1200.....E}
}

@ARTICLE{HoffmanGelman2011,
       author = {{Hoffman}, Matthew D. and {Gelman}, Andrew},
        title = "{The No-U-Turn Sampler: Adaptively Setting Path Lengths in Hamiltonian Monte Carlo}",
      journal = {arXiv e-prints},
         year = 2011,
        month = nov,
          eid = {arXiv:1111.4246},
        pages = {arXiv:1111.4246},
          doi = {10.48550/arXiv.1111.4246},
archivePrefix = {arXiv},
       eprint = {1111.4246},
 primaryClass = {stat.CO},
       adsurl = {https://ui.adsabs.harvard.edu/abs/2011arXiv1111.4246H}
}

@ARTICLE{Hogg2010,
       author = {{Hogg}, David W. and {Myers}, Adam D. and {Bovy}, Jo},
        title = "{Inferring the Eccentricity Distribution}",
      journal = {ApJ},
         year = 2010,
        month = dec,
       volume = {725},
       number = {2},
        pages = {2166-2175},
          doi = {10.1088/0004-637X/725/2/2166},
archivePrefix = {arXiv},
       eprint = {1008.4146},
 primaryClass = {astro-ph.SR},
       adsurl = {https://ui.adsabs.harvard.edu/abs/2010ApJ...725.2166H}
}

@misc{Huber2017,
       author = {{Huber}, Daniel},
        title = "{isoclassify: v1.2}",
         year = 2017,
        month = may,
          eid = {10.5281/zenodo.573372},
          doi = {10.5281/zenodo.573372},
      version = {v1.2},
    publisher = {Zenodo},
       adsurl = {https://ui.adsabs.harvard.edu/abs/2017zndo....573372H}
}

@ARTICLE{IdaLin2004-metallicity,
       author = {{Ida}, Shigeru and {Lin}, D.~N.~C.},
        title = "{Toward a Deterministic Model of Planetary Formation. II. The Formation and Retention of Gas Giant Planets around Stars with a Range of Metallicities}",
      journal = {\apj},
         year = 2004,
        month = nov,
       volume = {616},
       number = {1},
        pages = {567-572},
          doi = {10.1086/424830},
archivePrefix = {arXiv},
       eprint = {astro-ph/0408019},
 primaryClass = {astro-ph},
       adsurl = {https://ui.adsabs.harvard.edu/abs/2004ApJ...616..567I}
}

@ARTICLE{Ida2013-eccentricity,
       author = {{Ida}, S. and {Lin}, D.~N.~C. and {Nagasawa}, M.},
        title = "{Toward a Deterministic Model of Planetary Formation. VII. Eccentricity Distribution of Gas Giants}",
      journal = {\apj},
         year = 2013,
        month = sep,
       volume = {775},
       number = {1},
          eid = {42},
        pages = {42},
          doi = {10.1088/0004-637X/775/1/42},
archivePrefix = {arXiv},
       eprint = {1307.6450},
 primaryClass = {astro-ph.EP},
       adsurl = {https://ui.adsabs.harvard.edu/abs/2013ApJ...775...42I}
}

@ARTICLE{Jappsen2005,
       author = {{Jappsen}, A. -K. and {Klessen}, R.~S. and {Larson}, R.~B. and {Li}, Y. and {Mac Low}, M. -M.},
        title = "{The stellar mass spectrum from non-isothermal gravoturbulent fragmentation}",
      journal = {\aap},
         year = 2005,
        month = may,
       volume = {435},
       number = {2},
        pages = {611-623},
          doi = {10.1051/0004-6361:20042178},
archivePrefix = {arXiv},
       eprint = {astro-ph/0410351},
 primaryClass = {astro-ph},
       adsurl = {https://ui.adsabs.harvard.edu/abs/2005A&A...435..611J}
}

@ARTICLE{JuricTremaine2008,
       author = {{Juri{\'c}}, Mario and {Tremaine}, Scott},
        title = "{Dynamical Origin of Extrasolar Planet Eccentricity Distribution}",
      journal = {\apj},
         year = 2008,
        month = oct,
       volume = {686},
       number = {1},
        pages = {603-620},
          doi = {10.1086/590047},
archivePrefix = {arXiv},
       eprint = {astro-ph/0703160},
 primaryClass = {astro-ph},
       adsurl = {https://ui.adsabs.harvard.edu/abs/2008ApJ...686..603J}
}

@ARTICLE{Kipping2013,
       author = {{Kipping}, D.~M.},
        title = "{Parametrizing the exoplanet eccentricity distribution with the beta  distribution.}",
      journal = {\mnras},
         year = 2013,
        month = jul,
       volume = {434},
        pages = {L51-L55},
          doi = {10.1093/mnrasl/slt075},
archivePrefix = {arXiv},
       eprint = {1306.4982},
 primaryClass = {astro-ph.EP},
       adsurl = {https://ui.adsabs.harvard.edu/abs/2013MNRAS.434L..51K}
}

@ARTICLE{MaGe2014,
       author = {{Ma}, Bo and {Ge}, Jian},
        title = "{Statistical properties of brown dwarf companions: implications for different formation mechanisms}",
      journal = {\mnras},
         year = 2014,
        month = apr,
       volume = {439},
       number = {3},
        pages = {2781-2789},
          doi = {10.1093/mnras/stu134},
archivePrefix = {arXiv},
       eprint = {1303.6442},
 primaryClass = {astro-ph.EP},
       adsurl = {https://ui.adsabs.harvard.edu/abs/2014MNRAS.439.2781M}
}

@ARTICLE{MaldonadoVillaver2017,
       author = {{Maldonado}, J. and {Villaver}, E.},
        title = "{Searching for chemical signatures of brown dwarf formation}",
      journal = {\aap},
         year = 2017,
        month = jun,
       volume = {602},
          eid = {A38},
        pages = {A38},
          doi = {10.1051/0004-6361/201630120},
archivePrefix = {arXiv},
       eprint = {1702.02904},
 primaryClass = {astro-ph.SR},
       adsurl = {https://ui.adsabs.harvard.edu/abs/2017A&A...602A..38M}
}

@ARTICLE{Masuda2022,
       author = {{Masuda}, Kento and {Petigura}, Erik A. and {Hall}, Oliver J.},
        title = "{Inferring the rotation period distribution of stars from their projected rotation velocities and radii: Application to late-F/early-G Kepler stars}",
      journal = {MNRAS},
         year = 2022,
        month = mar,
       volume = {510},
       number = {4},
        pages = {5623-5638},
          doi = {10.1093/mnras/stab3650},
archivePrefix = {arXiv},
       eprint = {2112.07162},
 primaryClass = {astro-ph.SR},
       adsurl = {https://ui.adsabs.harvard.edu/abs/2022MNRAS.510.5623M}
}

@ARTICLE{MataSanchez2014,
       author = {{Mata S{\'a}nchez}, D. and {Gonz{\'a}lez Hern{\'a}ndez}, J.~I. and {Israelian}, G. and {Santos}, N.~C. and {Sahlmann}, J. and {Udry}, S.},
        title = "{Chemical abundances of stars with brown-dwarf companions}",
      journal = {\aap},
         year = 2014,
        month = jun,
       volume = {566},
          eid = {A83},
        pages = {A83},
          doi = {10.1051/0004-6361/201423803},
archivePrefix = {arXiv},
       eprint = {1404.0824},
 primaryClass = {astro-ph.EP},
       adsurl = {https://ui.adsabs.harvard.edu/abs/2014A&A...566A..83M}
}

@ARTICLE{McKeeOstriker2007,
       author = {{McKee}, Christopher F. and {Ostriker}, Eve C.},
        title = "{Theory of Star Formation}",
      journal = {\araa},
         year = 2007,
        month = sep,
       volume = {45},
       number = {1},
        pages = {565-687},
          doi = {10.1146/annurev.astro.45.051806.110602},
archivePrefix = {arXiv},
       eprint = {0707.3514},
 primaryClass = {astro-ph},
       adsurl = {https://ui.adsabs.harvard.edu/abs/2007ARA&A..45..565M}
}

@ARTICLE{Nagpal2023,
       author = {{Nagpal}, Vighnesh and {Blunt}, Sarah and {Bowler}, Brendan P. and {Dupuy}, Trent J. and {Nielsen}, Eric L. and {Wang}, Jason J.},
        title = "{The Impact of Bayesian Hyperpriors on the Population-level Eccentricity Distribution of Imaged Planets}",
      journal = {\aj},
         year = 2023,
        month = feb,
       volume = {165},
       number = {2},
          eid = {32},
        pages = {32},
          doi = {10.3847/1538-3881/ac9fd2},
archivePrefix = {arXiv},
       eprint = {2211.02121},
 primaryClass = {astro-ph.EP},
       adsurl = {https://ui.adsabs.harvard.edu/abs/2023AJ....165...32N}
}

@ARTICLE{numpy:2020,
  author  = {Harris, Charles R. and Millman, K. Jarrod and
            van der Walt, Stéfan J and Gommers, Ralf and
            Virtanen, Pauli and Cournapeau, David and
            Wieser, Eric and Taylor, Julian and Berg, Sebastian and
            Smith, Nathaniel J. and Kern, Robert and Picus, Matti and
            Hoyer, Stephan and van Kerkwijk, Marten H. and
            Brett, Matthew and Haldane, Allan and
            Fernández del Río, Jaime and Wiebe, Mark and
            Peterson, Pearu and Gérard-Marchant, Pierre and
            Sheppard, Kevin and Reddy, Tyler and Weckesser, Warren and
            Abbasi, Hameer and Gohlke, Christoph and
            Oliphant, Travis E.},
  title   = {Array programming with {NumPy}},
  journal = {Nature},
  year    = {2020},
  volume  = {585},
  pages   = {357–362},
  doi     = {10.1038/s41586-020-2649-2}
}

@PHDTHESIS{Petigura2015,
       author = {{Petigura}, Erik Ardeshir},
        title = "{Prevalence of Earth-size Planets Orbiting Sun-like Stars}",
       school = {University of California, Berkeley},
         year = 2015,
        month = jan,
       adsurl = {https://ui.adsabs.harvard.edu/abs/2015PhDT........82P}
}

@ARTICLE{Pollack1996,
       author = {{Pollack}, James B. and {Hubickyj}, Olenka and {Bodenheimer}, Peter and {Lissauer}, Jack J. and {Podolak}, Morris and {Greenzweig}, Yuval},
        title = "{Formation of the Giant Planets by Concurrent Accretion of Solids and Gas}",
      journal = {\icarus},
         year = 1996,
        month = nov,
       volume = {124},
       number = {1},
        pages = {62-85},
          doi = {10.1006/icar.1996.0190},
       adsurl = {https://ui.adsabs.harvard.edu/abs/1996Icar..124...62P}
}

@ARTICLE{pymc3,
       author = {{Salvatier}, John and {Wiecki}, Thomas and {Fonnesbeck}, Christopher},
        title = "{Probabilistic Programming in Python using PyMC}",
      journal = {arXiv e-prints},
         year = 2015,
        month = jul,
          eid = {arXiv:1507.08050},
        pages = {arXiv:1507.08050},
          doi = {10.48550/arXiv.1507.08050},
archivePrefix = {arXiv},
       eprint = {1507.08050},
 primaryClass = {stat.CO},
       adsurl = {https://ui.adsabs.harvard.edu/abs/2015arXiv150708050S}
}

@ARTICLE{Rosenthal2021,
       author = {{Rosenthal}, Lee J. and {Fulton}, Benjamin J. and {Hirsch}, Lea A. and {Isaacson}, Howard T. and {Howard}, Andrew W. and {Dedrick}, Cayla M. and {Sherstyuk}, Ilya A. and {Blunt}, Sarah C. and {Petigura}, Erik A. and {Knutson}, Heather A. and {Behmard}, Aida and {Chontos}, Ashley and {Crepp}, Justin R. and {Crossfield}, Ian J.~M. and {Dalba}, Paul A. and {Fischer}, Debra A. and {Henry}, Gregory W. and {Kane}, Stephen R. and {Kosiarek}, Molly and {Marcy}, Geoffrey W. and {Rubenzahl}, Ryan A. and {Weiss}, Lauren M. and {Wright}, Jason T.},
        title = "{The California Legacy Survey. I. A Catalog of 178 Planets from Precision Radial Velocity Monitoring of 719 Nearby Stars over Three Decades}",
      journal = {\apjs},
         year = 2021,
        month = jul,
       volume = {255},
       number = {1},
          eid = {8},
        pages = {8},
          doi = {10.3847/1538-4365/abe23c},
archivePrefix = {arXiv},
       eprint = {2105.11583},
 primaryClass = {astro-ph.EP},
       adsurl = {https://ui.adsabs.harvard.edu/abs/2021ApJS..255....8R}
}

@ARTICLE{Rosenthal2022,
       author = {{Rosenthal}, Lee J. and {Knutson}, Heather A. and {Chachan}, Yayaati and {Dai}, Fei and {Howard}, Andrew W. and {Fulton}, Benjamin J. and {Chontos}, Ashley and {Crepp}, Justin R. and {Dalba}, Paul A. and {Henry}, Gregory W. and {Kane}, Stephen R. and {Petigura}, Erik A. and {Weiss}, Lauren M. and {Wright}, Jason T.},
        title = "{The California Legacy Survey. III. On the Shoulders of (Some) Giants: The Relationship between Inner Small Planets and Outer Massive Planets}",
      journal = {\apjs},
         year = 2022,
        month = sep,
       volume = {262},
       number = {1},
          eid = {1},
        pages = {1},
          doi = {10.3847/1538-4365/ac7230},
archivePrefix = {arXiv},
       eprint = {2112.03399},
 primaryClass = {astro-ph.EP},
       adsurl = {https://ui.adsabs.harvard.edu/abs/2022ApJS..262....1R}
}

@ARTICLE{Santos2017,
       author = {{Santos}, N.~C. and {Adibekyan}, V. and {Figueira}, P. and {Andreasen}, D.~T. and {Barros}, S.~C.~C. and {Delgado-Mena}, E. and {Demangeon}, O. and {Faria}, J.~P. and {Oshagh}, M. and {Sousa}, S.~G. and {Viana}, P.~T.~P. and {Ferreira}, A.~C.~S.},
        title = "{Observational evidence for two distinct giant planet populations}",
      journal = {\aap},
         year = 2017,
        month = jul,
       volume = {603},
          eid = {A30},
        pages = {A30},
          doi = {10.1051/0004-6361/201730761},
archivePrefix = {arXiv},
       eprint = {1705.06090},
 primaryClass = {astro-ph.EP},
       adsurl = {https://ui.adsabs.harvard.edu/abs/2017A&A...603A..30S}
}
\bibliographystyle{aasjournal}

\end{document}